\documentstyle[preprint,aps,prb]{revtex}
\tightenlines
\begin{document}
\newcommand{\Jga}{{\it J}$_{Ga}$}
\newcommand{\Jsc}{{\it J}$_{Sc}$}
\newcommand{\Jn}{{\it J}$_{N}$}

\title{ScGaN Alloy Growth by Molecular Beam Epitaxy: Evidence for a Metastable Layered Hexagonal Phase}

\author{Costel Constantin, Hamad Al-Brithen, Muhammad B. Haider, David Ingram, and Arthur R.
Smith\footnote{Corresponding author, smitha2@ohio.edu}\\
\small{Condensed Matter and Surface Science Program, Department of
Physics and Astronomy, Ohio University, Athens, OH 45701}\\ }
\maketitle

\begin{abstract}
Alloy formation in ScGaN is explored using rf molecular beam epitaxy over the Sc fraction range $x$ = 0-100\%. Optical
and structural analysis show separate regimes of growth, namely I) wurtzite-like but having local lattice distortions
in the vicinity of the Sc$_{Ga}$ substitutions for small $x$ ($x$ $\leq$ 0.17), II) a transitional regime for
intermediate $x$, and III) cubic, rocksalt-like for large $x$ ($x$ $\geq$ 0.54). In regimes I and III, the direct
optical transition decreases approximately linearly with increasing $x$ but with an offset over region II. Importantly,
it is found that for regime I, an anisotropic lattice expansion occurs with increasing $x$ in which $a$ increases much
more than $c$. These observations support the prediction of Farrer and Bellaiche [Phys. Rev. B {\bf 66}, 201203-1
(2002)] of a metastable layered hexagonal phase of ScN, denoted $h$-ScN.
\end{abstract}

\vspace{1 cm}

\begin{center}{\bf PACS:} 78.66.Fd; 61.14.Hg; 61.66.Dk; 68.55.Nq; 61.10.Nz

\end{center}

ScN is a rocksalt (cubic) semiconductor with an indirect bandgap from $\Gamma$ $\rightarrow$ X of $\sim$ 1 eV and a
direct transition $E_t$ at the X point of 2.1-2.4 eV.\cite{Dismukes2,Gall,Gall2,Moustakas,Smith,Brithen3} It is
interesting to investigate the possibility of nitride bandgap engineering by putting Sc into GaN.\cite{Dismukes}
Takeuchi predicted a metastable wurtzite phase ($w$-ScN) for ScN.\cite{Takeuchi} However, recently Farrer and Bellaiche
have found that $w$-ScN should be unstable and that instead a layered hexagonal phase having nearly 5-fold
coordination, denoted {\it h}-ScN, which can be arrived at by flattening the bilayer of the wurtzite structure, should
be metastable.\cite{Farrer,Ranjan} In fact, such a structure has also been predicted to exist for MgO
($h$-MgO).\cite{Lambrecht} Yet, aside from the report of Little \& Kordesch, who grew ScGaN by sputtering and reported
amorphous or microcrystalline films and a linearly decreasing optical band gap (from 3.5 down to 2.0 eV) with
increasing Sc concentration,\cite{Kordesch} little is known experimentally regarding the actual crystal structures of
ScGaN alloys. For example, one might expect more than one growth regime for crystalline films. If so, then novel
transitional properties, as suggested by Farrer and Bellaiche, might be observed at the boundary between the two
different regimes.\cite{Farrer}

In this Rapid Communication, we present results for ScGaN film growth using a custom molecular beam epitaxy (MBE)
system that employs Ga and Sc effusion cells and a rf plasma N source. The sapphire(0001) substrates are first
nitridated at 650 $^\circ$C for 15 minutes with N plasma source operating at 500 W with a N$_{2}$ flow rate of 1.1 sccm
($P_{chamber}$ = $9\times 10^{-6}$ Torr). The ScGaN is then grown at specific flux ratio $r$ = \Jsc/(\Jsc+\Jga) at 650
$^\circ$C to a thickness of 250-340 nm. The N flux is held constant with growth in N-rich conditions (\Jsc + \Jga $<$
\Jn). The growth is monitored {\it in-situ} by reflection high energy electron diffraction (RHEED) using a 20 keV
e$^-$-beam, and the films are studied {\it ex-situ} by X-ray diffraction (XRD) with Cu {\it K}$\alpha$ X-rays, optical
absorption (OA), and Rutherford backscattering (RBS).

A series of ScGaN films on sapphire(0001) with $r$ in the range between 0 and 1 have been grown and analyzed. Presented
in Fig. 1 are the RHEED patterns of 8 such ScGaN films, including the 2 endpoints at r=0 [GaN(000$\bar{1}$)] and r=1
[ScN(111)]. The direction of the RHEED beam is along [11$\bar{2}$0] for GaN and correspondingly [1$\bar{1}$0] for ScN.

Calculation of the RHEED patterns is accomplished by considering the structural models. Figure 2(a) shows a side-view
model of the [000$\bar{1}$]-oriented wurtzite GaN bi-layers as viewed along [11$\bar{2}$0], and Fig. 2(b) shows the
associated reciprocal space map with spot indices $\nu_1$ $\nu_2$ $\nu_3$. The intensity of the diffraction spots with
$\nu_1$ odd, calculated from the structure factor for wurtzite, alternates along the $k_z$ direction, and the spots
with $\nu_1$ = 0 and $\nu_3$ odd have zero intensity. Clearly the RHEED pattern of GaN(000$\bar{1}$) shown in Fig. 1(a)
is in good agreement with the calculated reciprocal space map, and the measured ratio of $S_{in,GaN}$ to $S_{out,GaN}$
for GaN is 1.92, which agrees well with the expected value $c$/($\sqrt{3}a$/2) = 5.185 \AA/2.762 \AA\ = 1.877.

Fig. 2(c) shows the side-view model of ScN(111) as viewed along [1$\bar{1}$0] together with the corresponding
reciprocal space map shown in Fig. 2(d). The spot intensities are calculated from a superposition of the structure
factors of two inequivalent 111-oriented fcc grains (note that fcc is not 2-fold symmetric about 111, and for a single
grain the diffraction pattern is asymmetrical about the 00 rod). Clearly the RHEED pattern of ScN(111) shown in Fig.
1(e) is in excellent agreement with the calculated reciprocal space map in Fig. 2(d), and the ratio of the lateral spot
spacing S$_{in,ScN}$ to S$_{out1,ScN}$ is measured to be 2.86, in good agreement with the expected value of 2$\sqrt{2}$
$\simeq$ 2.83.

Comparing the rocksalt and wurtzite patterns, we note that the measured spacing S$_{out2,ScN}$ [see Fig. 1(e)] is very
close to 2$\times$ the measured spacing S$_{out,GaN}$ [see Fig. 1(a)], due to the fact that the Sc layer spacing
$a/\sqrt{3}$ = 2.599 \AA\ in ScN is very close to the Ga layer spacing $c/2$ = 2.592 \AA\ in GaN. Yet clearly the RHEED
patterns of wurtzite and rocksalt are easily distinguishable.

Shown in Figs. 1(b-d) are the RHEED patterns for ScGaN alloy films having $r$ in the range 0$<$r$\leq$0.29. These RHEED
patterns are in good agreement with the wurtzite reciprocal space map. However, we note that the structure factor for
the layered hexagonal phase is identical to that of wurtzite. To determine if these ScGaN samples are consistent with
wurtzite or not, it is necessary to consider the measured lattice constants $a$ and $c$ (discussed below).

Shown in Figs. 1(f-h) are the RHEED patterns for ScGaN films having $r$ in the range 0.54$\leq$r$<$1. Although each
pattern shows a degree of polycrystallinity (based on the ring-like RHEED pattern), each pattern is in good agreement
with the rocksalt-type pattern shown in Figs. 1(e) and 2(d). The results suggest that each monocrystal has rocksalt
structure for growth at large Sc composition, consistent with predictions.\cite{Farrer,Takeuchi2}

Incorporation $x$ $\equiv$ $N_{Sc}$/($N_{Sc}$ + $N_{Ga}$), where $N_{Sc}$ and $N_{Ga}$ are the number of Sc and Ga
atoms within a given volume respectively, was obtained using RBS; the results are given in Table I. Whereas $r$ and $x$
are about the same for the cubic regime ($r$ $\geq$ 0.54), for the hexagonal regime ($r$ $\leq$ 0.29), $x$ is
consistently smaller than $r$, suggesting that the sticking coefficients $S_{Ga}$ and $S_{Sc}$ are different at low $r$
values but similar at high $r$ values. This behavior is reasonable given that the N-polar (and N-terminated)
wurtzite-like surface would have a single dangling bond per N atom (film polarity is determined in a later experiment
by the RHEED pattern after subsequent growth of a GaN layer under Ga-rich conditions)\cite{SmithAPL98,SmithPRL}. In
comparison, from the model of Fig. 2(c), the (111)-oriented, N-terminated rocksalt surface would have 3 dangling bonds
per N atom, resulting in larger surface diffusion barrier of both Sc and Ga atoms, rougher growth mode, and larger
sticking coefficients for both Sc and Ga. This agrees with the RHEED patterns which are more spotty for the cubic
regime compared to the hexagonal regime.

The lattice constant $a$ is directly obtained from the RHEED pattern using a peak fitting program. The RHEED
calibration is performed using a GaN substrate grown by metal-organic chemical vapor deposition. The resulting $a$
values are plotted vs. $x$ in Fig. 4 for the small $x$ values, where it is seen that $a$ increases with $x$.

Crystallinity and {\it out-of-plane} lattice constant information was determined using XRD, and shown in Fig. 3 are
results for ScGaN layers with low $x$. Figure 3(a) shows the entire XRD spectrum (20-140$^\circ$) for the film with $x$
= 0.05 where 5 peaks are seen - 2 sapphire peaks (0006 and 00012) and 3 ScGaN peaks (0002, 0004, and 0006). The spectra
are corrected so that both sapphire peaks give the same lattice constant ($c$ = 12.98 \AA); the alignment of the
sapphire 0006 peaks for films with $x$ = 0.05, 0.14, and 0.17 is presented in Figure 3(b). The observed ScGaN peak
shift for the same 3 samples is shown in Fig. 3(c); compared to $x$ = 0, the 0002 peak shifts to the left for $x$ =
0.05 and 0.14 but interestingly shifts slightly back to the right for $x$ = 0.17, from which the $c$-spacings are
calculated using Gaussian peak fitting and Bragg's law to be 5.190 \AA, 5.195 \AA, and 5.194 \AA, respectively,
compared to 5.188 \AA\ for $x$ = 0. The $c$ values are also plotted in Fig. 4 vs. $x$.

Over the range 0$<$x$<$0.17, $a$ increases by a net 0.08 \AA\ while $c$ increases by only a net 0.006 \AA. Thus for the
small $x$ range, the Sc$_x$Ga$_{1-x}$N lattice expands predominantly within the $c$-plane. Such anisotropic expansion
as well as the decrease of $c$ between $x$ = 0.14 and 0.17 would not be expected in the case of alloying of
iso-crystalline binary compounds. The anisotropic expansion suggests the following picture, as illustrated by the
schematic model shown in Fig. 2(e): at the location where a Sc atom substitutes for a Ga atom, a local lattice
distortion occurs in which the N-Sc-N bond angle $\theta_b$ (having one leg in the $c$-direction) reduces compared to
the $\sim$ 108$^\circ$ for a wurtzite bond (in ideal $h$-ScN, $\theta_b$ = 90$^\circ$). In other words, the internal
parameter $u$ [see Fig. 2(a)] varies within the alloy away from the $w$-GaN value 0.376.\cite{Farrer} For $h$-ScN ($x$
= 1), Farrer {\it et al.} have predicted $a$ = 3.66 \AA, $c$ = 4.42 \AA, and $c$/$a$ = 1.207.\cite{Farrer} Thus $a$ can
potentially locally increase by up to 14.77\% of $a_{GaN}$ = 3.189 \AA, depending on $x$; at the same time, $c$ would
locally decrease.

Further experimental evidence for the low $x$ regime points to the same model. First, we note substantial broadening of
the RHEED diffraction lines with increasing $x$ (see FWHM values vs. $x$ in Table I). Second, we note the substantial
intensity decrease and broadening of the 0002 ScGaN XRD peak with increasing $x$ [see Fig. 3(c) and also Table I]. Such
behavior is indicative of an increased spread of lattice constants with increasing $x$, resulting in reduction of the
long range order or of the maximum correlation length of the crystal.

Very recently, Ranjan and Bellaiche have calculated the $c/a$ ratio for an ideally ordered Sc$_x$Ga$_{1-x}$N having $x$
= 0.5 to be 1.55.\cite{Bellaiche}. Considering the $c/a$ ratio for GaN ($x$ = 0) to be 1.626 = 5.185 \AA/3.189 \AA,
linear interpolation gives decreasing values of $c/a$ for increasing $x$. This data is presented in Table 1 which also
shows values for the ideal wurtzite $c/a$ ratios vs. $x$ (based on \cite{Takeuchi2}). The data show that the change in
$c/a$ over the range $x$ = 0 to 0.17 is -0.01 for the wurtzite case versus -0.03 for the model of Ranjan and Bellaiche.
The experiment, which finds a change in $c/a$ over this range to be -0.040, is thus in better agreement with the
latter.

Extrapolating straight line fits of the measured $a$ and $c$ vs. $x$ data to $x$ = 1 results in $a$($x$=1) = 3.60 \AA\
and $c$/$a$ = 1.45 for a hypothetical ScN in hexagonal phase. Thus the extrapolated $c$/$a$ ratio is significantly
smaller than that predicted for $w$-ScN (1.6); and, the extrapolated $a$ is significantly larger than that predicted
for $w$-ScN (3.49 \AA),\cite{Takeuchi} closer in fact to the value predicted for $h$-ScN (3.66 \AA).\cite{Farrer} In
fact, we note that $a$ increases faster, and that $c$ decreases, with $x$ between 0.14 and 0.17.

Optical absorption measurements have been obtained for all the films. Shown in Fig. 5(a) is plotted the quantity
($h\nu\alpha$)$^{2}$ vs. $h\nu$ for each film. The $E_t$ is estimated from the energy intercept of a straight line
tangent to the curve near its inflection point. The resulting $E_t$ values are plotted vs. $r$ and $x$ in Fig. 5(b).
The $E_t$ = 3.37 eV of wurtzite GaN is obtained at $x$ = 0. The $E_t$ = 2.15 eV is obtained at $r$ = 1 for ScN grown on
MgO(001).\cite{Smith} As can be seen, three different regions, consistent with the RHEED results, can be distinguished
: I) low Sc fraction (0$<r<$0.30); II) intermediate Sc fraction - transitional regime (0.30$<r<$0.54); and III) high Sc
fraction (0.54$<r<$1). In both regions I and III, $E_t$ decreases monotonically with increasing $x$.

For region I, extrapolating a straight line fit of the $E_t$ values to $x$ = 1 (pure ScN) obtains an $E_t$ = 2.3 eV -
significantly smaller than the value of $\sim$ 3.0 eV predicted for the $E_t$ of $w$-ScN.\cite{Takeuchi} The
extrapolation from low $x$ however is probably not a good estimate of the $E_t$ of $h$-ScN since layered hexagonal is
not iso-crystalline with the low $x$ regime (wurtzite with local N-Sc-N bond distortions).

As RHEED indicates, rocksalt structure is observed for larger $x$ (region III), and within region III, $E_t$ decreases
linearly towards the rocksalt value of 2.15 eV at $x$ = 1. Using a straight line fit to the $E_t$ values of region III
and extrapolating to $x$ = 0 yields an $E_t$ = 2.72 eV for a hypothetical GaN in rocksalt structure. Finally, the $E_t$
values in region II (encircled points) with $r$ in the range 0.30-0.51 show comparitively large variations, which are
consistent with the expected instability of the crystal structure near the transition between the stable hexagonal and
cubic regimes.

To summarize, we have shown that the disparate ground state crystal structures of ScN and GaN lead to two distinct
regimes of structural and optical properties. For both low $x$ and high $x$, alloy-type behavior is observed. For $x$
$\geq$ 0.54, rocksalt structure is found, in agreement with predictions.\cite{Farrer,Takeuchi2} For small $x$ up to
0.17, an anisotropic expansion of the ScGaN lattice is observed which is interpreted in terms of local lattice
distortions of the wurtzite structure in the vicinity of Sc$_{Ga}$ substitutional sites in which there is a decrease of
the N-Sc-N bond angle. This tendency toward flattening of the wurtzite bilayer is consistent with a predicted $h$-ScN
phase.\cite{Farrer}

This work is supported by the National Science Foundation under Grant No. 9983816.

\begin{figure}
\caption {(a) RHEED pattern of wurtzite GaN(000$\bar{1}$) along [11$\bar{2}$0]; (b)-(d) RHEED patterns of
Sc$_x$Ga$_{1-x}$N for low Sc concentration along same azimuth as (a); (e) RHEED pattern of rocksalt ScN (111) along
[1$\bar{1}$0]; (f)-(h) RHEED pattern of Sc$_x$Ga$_{1-x}$N for high Sc concentration along same azimuth as (e).}
\end{figure}

\begin{figure}
\caption{(a) Side view diagram of GaN [000$\bar{1}$] along [11$\bar{2}$0]; (b) reciprocal space map corresponding to
(a), 3-index notation corresponds to $\nu_1$ $\nu_2$ $\nu_3$ which label the reciprocal-lattice points; (c) side view
diagram of ScN [111] along [1$\bar{1}$0]; (d) reciprocal space map corresponding to (c) using similar 3-index notation.
Size of dots is proportional to their intensity; (e) schematic model of ScGaN for low $x$ regime showing local
distortions of the bond angle $\theta_b$.}
\end{figure}

 \begin{figure}
\caption {(a) XRD of a representative spectrum with $x$ = 0.05 showing the two sapphire peaks [0006 \& 00012] and three
ScGaN peaks [0002, 0004, \& 0006]; (b) sapphire 0006 peaks for the films with $x$ = 0.05; 0.14; 0.17; (c) ScGaN 0002
peaks at 2$\theta$ $\sim$ 34.5$^\circ$ for the films with $x$ = 0.05; 0.14; 0.17.}
\end{figure}

\begin{figure}
\caption {The lattice spacings $a$ and $c$ vs. $x$ for low Sc concentration.}
\end{figure}

\begin{figure}
\caption {(a) Optical absorption measurements of all the films; numbers are the Sc/(Sc + Ga) flux ratios $r$, and
tangent line for r = 0.89 exemplifies method of obtaining $E_t$. Small peaks at $\sim$ 3.4 eV for large $x$ are
instrumental in origin and of no importance here. (b) deduced $E_t$ values versus $r$ and $x$ for all the films showing
the 3 different regions. Most error bars are too small to be seen.}
\end{figure}

\newpage
\begin{table}[ht]
\caption{Flux ratio $r$ = $J_{Sc}$/($J_{Sc}$ + $J_{Ga}$), Sc composition $x$ = $N_{Sc}$/($N_{Sc}$ + $N_{Ga}$), XRD 0002
peak amplitude, XRD 0002 FWHM, RHEED FWHM, measured $c/a$ ratio, expected $c/a$ ratio (based on Ref. \cite{Bellaiche}),
and $c/a$ ratio for ideal wurtzite structure (based on Ref.\cite{Takeuchi2} for different Sc$_x$Ga$_{1-x}$N samples).
Sample 1-4 correspond to regime I, samples 5-7 to regime III.}

\begin{tabular}{ccccc|ccc}

ScGaN Sample  & 1 & 2 & 3 & 4 & 5 & 6 & 7 \\ \hline

flux ratio $r$   & 0 & 0.068  & 0.21 & 0.29 & 0.54 & 0.78 & 0.89\\

composition $x$  & 0 & 0.05 & 0.14 & 0.17 & 0.54 & 0.74 & 0.89\\

0002 XRD peak Ampl. & 82182 & 51443 & 13086 & 5438 & - & - & -\\

0002 XRD FWHM & 0.149$^\circ$ & 0.181$^\circ$ & 0.277$^\circ$ & 0.349$^\circ$ & - & - & -\\

RHEED FWHM & 8.1 & 9.7 & 19.4 & 22.6 & - & - & -\\

measured $c/a$  & 1.635 & 1.621 & 1.612 & 1.595 & - & - & -\\

$c/a$ expected based on Ref. \cite{Bellaiche} & 1.626 & 1.62 & 1.60 & 1.60 & - & - & -\\

ideal wurzite $c/a$ based on Ref. \cite{Takeuchi2} & 1.62 & 1.62 & 1.61 & 1.61 & - & - & -\\

\end{tabular}
\end{table}

\end{document}